\documentclass[aps,pra,twocolumn,groupedaddress]{revtex4-2}
\usepackage{graphicx}
\usepackage{siunitx}
\usepackage{hyperref}
\usepackage{amsmath}

\begin{document}


\title{Luminescence thermometry based on photon emitters in nanophotonic silicon waveguides}


\author{Kilian Sandholzer}
\email[]{kilian.sandholzer@tum.de}
\author{Stephan Rinner}
\author{Justus Edelmann}
\author{Andreas Reiserer}
\email[]{andreas.reiserer@tum.de}
\affiliation{Technical University of Munich, TUM School of Natural Sciences, Department of Physics and Munich Center for Quantum Science and Technology (MCQST), James-Franck-Straße 1, D-85748 Garching, Germany, and Max-Planck-Institute of Quantum Optics, Hans-Kopfermann-Straße, 1, D-85748 Garching, Germany.}


\date{\today}

\begin{abstract}
The reliable measurement and accurate control of the temperature within nanophotonic devices is a key prerequisite for their application in both classical and quantum technologies. Established approaches use sensors that are attached in proximity to the components, which only offers a limited spatial resolution and thus impedes the measurement of local heating effects. Here, we therefore study an alternative temperature sensing technique that is based on measuring the luminescence of erbium emitters directly integrated into nanophotonic silicon waveguides. To cover the entire temperature range from 295~K to 2~K, we investigate two different approaches: The thermal activation of non-radiative decay channels for temperatures above $\SI{200}{\kelvin}$ and the thermal depopulation of spin- and crystal field levels at lower temperatures. The achieved sensitivity is $0.22(4)\,\mathrm{\%/K}$ at room temperature and increases up to $420(50)\,\mathrm{\%/K}$ at approximately $\SI{2}{\kelvin}$. Within a few-minute measurement interval, we thus achieve a measurement precision that ranges from 0.04(1)~K at the lowest studied temperature to 6(1)~K at ambient conditions. In the future, the measurement time can be further reduced by optimizing the excitation pulse sequence and the fiber-to-chip coupling efficiency. Combining this with spatially selective implantation promises precise thermometry from ambient to cryogenic temperatures with a spatial resolution down to a few nanometers.
\end{abstract}


\maketitle
\section{Introduction}

Thermal management in nanophotonic devices is of utmost importance in many fields of research and technology, in particular with current developments towards smaller footprints with increasing numbers of components per chip \cite{karabchevsky_-chip_2020}, and in the field of quantum photonic technologies \cite{wang_integrated_2020,huang_trends_2020,gonzalez-tudela_lightmatter_2024}. One of the most mature platforms in these developments is silicon \cite{adcock_advances_2021, shekhar_roadmapping_2024}. Its various applications require accurate temperature control in a very broad range. As an example, thermally controlled switches \cite{song_fast_2008} or temperature-sensitive modulators \cite{manipatruni_wide_2008, reed_silicon_2010} operate at room temperature for optical transmission, classical \cite{zhou_silicon_2024} and quantum \cite{zhu_quantum_2024} information processing. On the other hand, integrated photon detectors on silicon perform better at low temperatures, including germanium diodes that show suppressed dark currents \cite{koester_temperature-dependent_2006} and single-photon detection capabilities at 125~K \cite{vines_high_2019}, as well as superconducting nanowire single-photon detectors that require cryogenic temperature \cite{shibata_waveguide-integrated_2019}. 
Finally, the generation of quantum light with single emitters in silicon \cite{redjem_single_2020,gritsch_purcell_2023} and the optical readout and control of spin qubits coupled to silicon photonic cavities \cite{gritsch_optical_2024} requires temperatures in the few-Kelvin regime for coherent operation. 

Controlling the temperature in such devices demands sensors that combine fast and accurate measurements with a compact design to minimize disturbance and allow for nanoscale spatial resolution sufficient to resolve individual components or even to measure temperature gradients within them. 
Here, we introduce a new technique that allows for such temperature measurements directly within nanophotonic silicon waveguides. Our approach uses the optical emission of erbium dopants embedded into the silicon host material \cite{gritsch_narrow_2022}.

In general, optical thermometers provide a convenient way to directly integrate sensors into nanophotonic structures with local read-out capabilities \cite{bradac_optical_2020}. 
Previous approaches in nanophotonic silicon devices were based on the thermo-optic effect on optical structures \cite{ma_progress_2020} such as resonators \cite{kim_silicon_2010,zhang_high_2016}, interferometers \cite{tao_demonstration_2015,chiang_highly_2020} and Bragg gratings \cite{Klimov_-chip_2015}. However, this has two disadvantages. First, these sensors are not integrated into the device under study, but require additional dedicated structures. Second, a high sensitivity cannot be achieved down to cryogenic temperature, as the thermo-optic effect in silicon is suppressed by four orders of magnitude at 5~K compared to 300~K \cite{komma_thermo-optic_2012}. 

Both difficulties can be avoided when using luminescence thermometry \cite{dedyulin_emerging_2022, brites_spotlight_2023} of light emitters such as color centers, lanthanide-based materials, and quantum dots \cite{harrington_luminescence_2024}. In particular, this approach provides a way to directly implement the thermal probes within nanophotonic devices. Still, previous experiments were often limited to specific temperature regimes and incompatible with silicon nanophotonic structures \cite{plakhotnik_luminescence_2010,acosta_temperature_2010,fujiwara_diamond_2021,alkahtani_tin-vacancy_2018,nguyen_all-optical_2018,fan_germanium-vacancy_2018, liu_all-optical_2024}. 
Yet, sensing across temperature ranges from \SI{4}{\kelvin} to room temperature has been reported for lanthanide-containing crystalline powders \cite{liu_mixed-lanthanoid_2015,ndala-louika_ratiometric_2017,brites_widening_2018,shang_dual-mode_2019,bolek_ga-modified_2021,li_investigation_2021,cai_high-sensitive_2024} with sensitivities of up to \SI{31}{\percent\per\kelvin} \cite{liu_mixed-lanthanoid_2015} and maximum precision of \SI{0.02}{\kelvin} \cite{bolek_ga-modified_2021}. In principle, these sensors can be reduced to the single nano- or microparticle level \cite{zhang_luminescence_2022} thus yielding high spatial resolution for the characterization of nanophotonic structures. However, their integration into nanophotonic silicon devices is challenging: Due to their operating principle, the required wavelengths for operation are in the absorption band of silicon, which entails heating and hinders direct integration into photonic devices. Furthermore, it necessitates free-space optical read-out, which would limit the compactness and footprint of such thermometers. Similarly, the recent approach of using erbium-chloride-silicate nanowires that operate above \SI{1500}{\nano\meter} \cite{liang_self-optimized_2023} would exhibit significant drawbacks in robustness, positioning, and thermal coupling to nanophotonic silicon structures.   

In this manuscript, we overcome these limitations by directly integrating erbium dopants into nanophotonic silicon waveguides, implementing temperature-dependent spectroscopy and using it to assess the performance of derived observables for thermometry. 
Using multiple temperature-dependent effects such as activation of non-radiative decays, depopulation of crystal field levels and electronic spin levels, we demonstrate precise thermal probes for the entire temperature range from 2~K to 300~K. 
Combined with the high degree of spatial control achievable in the erbium integration process, our method provides a reliable local temperature probe for silicon nanostructures. 

\section{Experimental setup}
\begin{figure}[ht]
\includegraphics[width=\columnwidth]{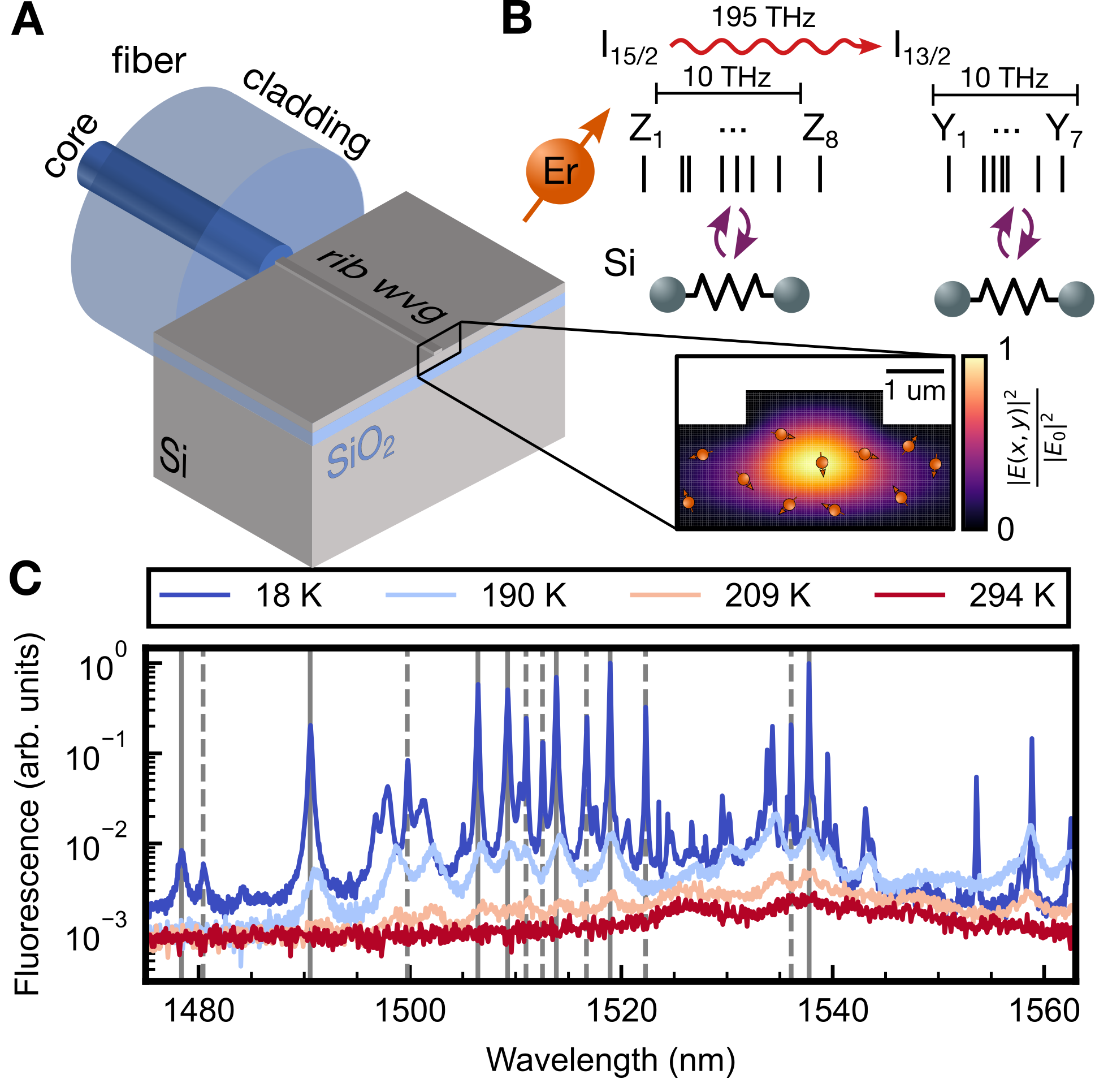} 
\caption{Erbium-based thermometry in nanophotonic silicon waveguides. (A) A cleaved single-mode fiber is glued to a rib waveguide fabricated on an erbium-doped silicon-on-insulator chip. The inset depicts the optical waveguide mode (red) in the homogeneously doped (orange spin symbols) device layer. (B) The crystal field levels in the ground- ($\text{I}_{15/2}$, $Z_1$ to $Z_8$) and optically excited ($\text{I}_{13/2}$, $Y_1$ to $Y_7$) manifold of the integrated erbium dopants thermalize (purple arrows) via phonons to the silicon host crystal. The emitters are excited by a tunable gated laser (red wiggly arrow) at telecommunications wavelengths (1475~nm - 1570~nm). (C) The photoluminescence signal after resonant excitation at different wavelengths depends on the temperature, shown at representative values between $\SI{18}{\kelvin}$ and $\SI{294}{\kelvin}$ (colors, top to bottom). Transitions between the level $Z_1$ to the $Y$ levels are indicated as vertical lines for the main integration sites A (solid) and B (dashed).\label{fig1}
}
\end{figure}

The experimental setup is sketched in Figure~\ref{fig1}A. A rib waveguide with 500~nm etch depth is fabricated using electron beam lithography on a silicon-on-insulator chip that has been homogeneously implanted with erbium ions and then annealed using the procedure described in earlier works \cite{gritsch_narrow_2022, rinner_erbium_2023}. The resulting simulated peak ion densities are between \SI{2d16}{\per\cubic\centi\metre} and \SI{4d16}{\per\cubic\centi\metre}. We investigate two samples that differ in the growth technique of the device layer. One is made from a crystal grown by the Float-Zone (FZ) technique, while the other one is grown by chemical vapor deposition (CVD). We find that both samples exhibit very similar spectra and properties, demonstrating that our technique is widely applicable and not restricted to a specific silicon host material.

A cleaved silica fiber is glued at the end of the chip to couple light in and out of the waveguide \cite{holzapfel_characterization_2024}. The chips are diced to a length of \SI{0.9}{\centi\meter} (FZ) and \SI{0.8}{\centi\meter} (CVD), which determines the approximate length of the probed waveguides. 
The intensity profile of the guided mode inside the waveguide is shown in the inset of Figure~\ref{fig1}A with an approximate cross-section of about \SI{1d-8}{\centi\square\meter}. 

The optical signal of the erbium dopants is dominated by two integration sites labeled A and B \cite{gritsch_narrow_2022}. 
Since the yield of these integration sites is less than \SI{1}{\percent}, we probe about \num{1d6} ions in total per waveguide.
The two lowest manifolds in the fine structure ($I_{15/2}$, $I_{13/2}$) define the optical transition frequency at $\nu_0\approx 195\ \mathrm{THz}$. They are further split by the crystal field (CF) as depicted for site A in Figure~\ref{fig1}B. The ground manifold consists of eight ($Z_1$ to $Z_8$), the excited manifold of seven levels ($Y_1$ to $Y_7$) that are Kramers doublets \cite{liu_electronic_2005} with a spread of the transitions frequency of about $20\ \mathrm{THz}$. The ground-to-ground transitions for sites A and B appear at \SI{1537.76}{\nano\meter} and \SI{1536.06}{\nano\meter}, respectively. The thermal occupation of the manifolds is given by Maxwell-Boltzmann statistics. As indicated in Figure~\ref{fig1}B, thermal equilibrium with the silicon host crystal is obtained via phononic coupling. From many earlier measurements in other rare-earth doped materials \cite{malkin_ion-phonon_2005}, the thermalization between different crystal field levels is expected to happen on a nanosecond timescale. Thus, both the ground- and optically excited states thermalize on a much shorter timescale than the optical decay, which exceeds \SI{10}{\micro\second} up to room temperature.

The erbium ensemble is probed via resonant photoluminescence excitation (PLE) with a narrowband laser using excitation pulses of $\SI{200}{\micro\second}$ duration and a typical power of around \SI{10}{\micro\watt} in the waveguide. Subsequently, fluorescence photons are detected using a superconducting nanowire single photon detector with photon counts per repetition of up to \num{130}. Probe and fluorescent light are coupled via the glued fiber that maintains its coupling efficiency of $\approx 4\,\%$ over the entire studied temperature range. 
The sample is glued to a mount made of oxygen-free copper and inserted into a closed-cycle helium cryostat to achieve temperature tuning in the range 295~K to 1.2~K. A calibrated resistive temperature sensor (Lake Shore Cernox) is thermally connected to the sample mount and used as a reference temperature probe. 
The optical transitions of sites A and B are distributed between 1475~nm and 1640~nm. We are mainly interested in the ground state transitions, which appear below 1538~nm. For some measurements, we thus use an optical long-pass filter with a  1.5~nm wide step centered at a wavelength of 1551.3~nm to increase the signal-to-noise ratio for sites A and B by blocking the emission from other sites with more narrow CF manifolds \cite{gritsch_narrow_2022}.

\section{Results and discussion}
The PLE spectra obtained after pulsed excitation exhibit a strong temperature dependence, as shown in Figure~\ref{fig1}C for representative datasets from cryogenic to room temperature. In these spectra, the mentioned long-pass filter is used to filter out background fluorescence. At room temperature, only a small signal is detected around the ground-to-ground-state transition ($Z_1$-$Y_1$), showing several features with a width of several nanometers. In this regime, the fluorescence is strongly reduced as non-radiative decays can occur via close-by impurities or defects \cite{coffa_temperature_1994,polman_erbium_1997,liu_quantification_2023}. Since this process relies on phononic coupling, an exponential increase of the signal is observed when lowering the temperature until about 200~K, where the optical decay starts to dominate.

The long-pass filter enhances this effect by blocking transitions from higher crystal field levels, which become thermally activated first, leading to the strong signal increase between the data for 209~K and 190~K. While the spectrum at 190~K shows already individual features of crystal field transitions of site A and B, the peaks are strongly broadened by phononic couplings between the crystal field levels. After further reducing the temperature to 18~K, the higher crystal field levels are almost completely depopulated, which shifts the mean fluorescence wavelength to longer wavelengths leading to an increase in signal transmission through the detection filter. The remaining transitions mainly originate from the $Z_1$ ground states of site A and B, which are indicated by vertical lines. They are getting more narrow with decreasing temperature and become individually resolved. Besides the integration sites A and B, additional sites also contribute to the spectrum \cite{gritsch_narrow_2022}. 

In conclusion, the spectrum exhibits several temperature-dependent parameters, and different thermometric probes will give optimal results depending on the temperature regime. This will be discussed in the following subsections.

\subsection{Photoluminescence quench regime\label{PLquench}}

As a first thermometric probe, we study the reduction of the fluorescence with increasing temperature. This exponential suppression of the overall signal when approaching room temperature, usually denoted as quench, has been detected previously for erbium-doped silicon in photoluminescence spectra after non-resonant excitation \cite{coffa_temperature_1994}. While the exact mechanism of the non-radiative decay is not fully understood, a quantitative description via the exponential activation of energy transfer to nearby defects has been reached \cite{liu_quantification_2023}. Here, we use this effect to derive a thermometric probe from an integrated amplitude response after excitation in a broad spectral range around $\SI{1535}{\nano\meter}$, where we observe the largest signal at room temperature.

\begin{figure}[ht]
\includegraphics[width=\columnwidth]{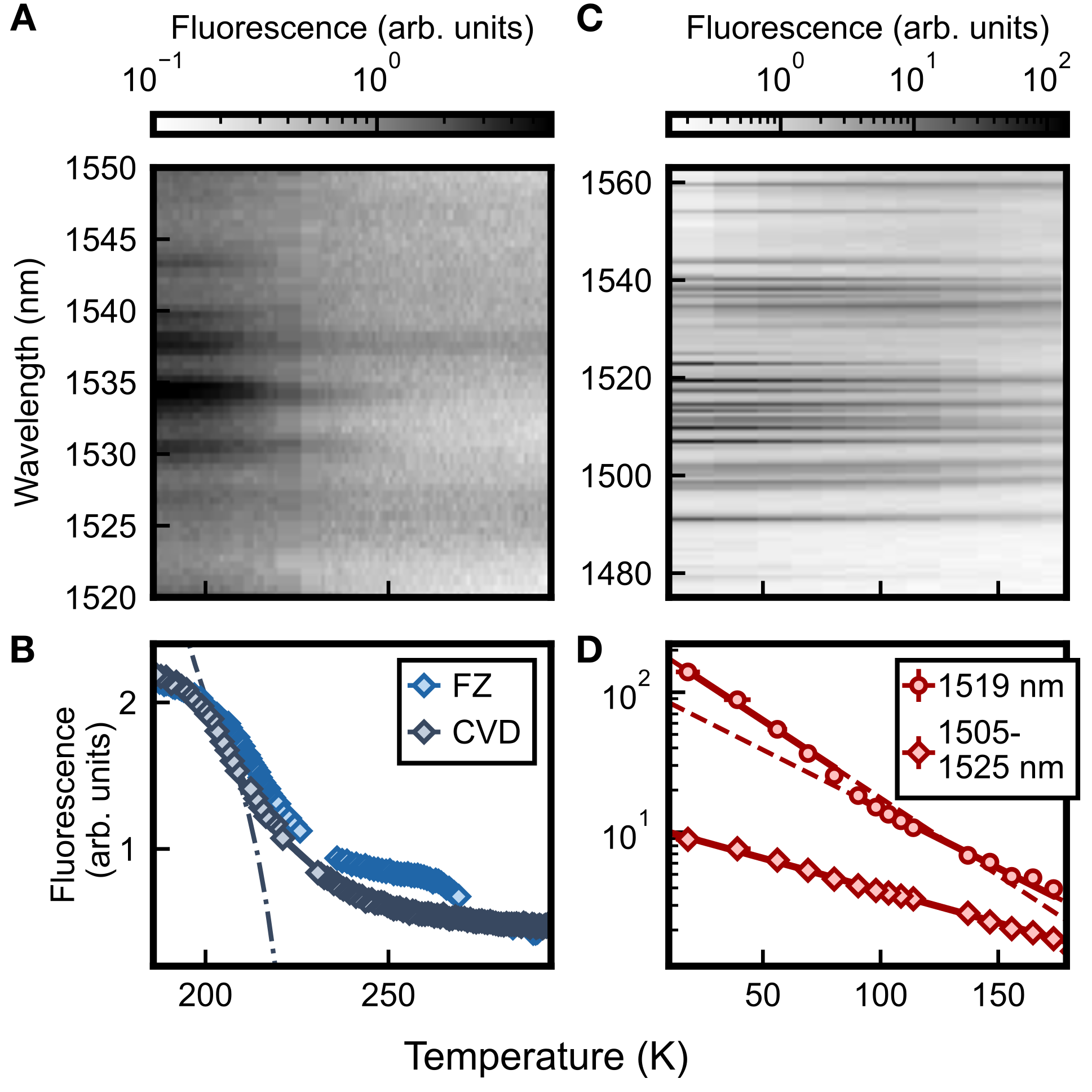}
\caption{Fluorescence after resonant excitation in different regimes. (A) In the photoluminescence quench regime, the fluorescence at all studied excitation wavelengths strongly decreases with temperature. (B) The average fluorescence in the range of 1520~nm to 1550~nm exhibits a similar temperature dependence when comparing two silicon samples fabricated by different growth techniques (FZ: float zone, CVD: chemical vapor deposition). (C) At lower temperatures, the relative height of the different lines is changed, as shown for the FZ sample in a long-pass filtered measurement. (D) The fluorescence at a wavelength of 1519~nm (red circles) and averaged over the range 1505~nm to 1525~nm (red diamonds) exhibits a strong temperature dependence. The solid lines in (B, D) indicate exponential fits to the data, while the dashed and dash-dotted lines show extensions beyond the fit range. The error bars denote the standard error of 500 averages.
\label{fig2}}
\end{figure}

In the quench regime, the fluorescence is rather faint. We thus study it without the long-pass filter, and use temporal gating to separate the emission from the excitation. In Figure~\ref{fig2}A, we show the wavelength dependence of the PLE spectrum, measured on the CVD sample in the temperature range from 180~K to 300~K. A quench of the emission is observed independent of the excitation wavelength. In the following, we thus average over all measured wavelengths to obtain the temperature dependence in Figure~\ref{fig2}B (grey diamonds) that follows an exponential decay at high temperatures (dashed fit curve) with a saturation behavior at lower temperatures as well captured by an exponential fit (dash-dotted fit curve). 
As the underlying decay process may depend strongly on defects of the host crystal, its temperature response could be different in different materials \cite{coffa_temperature_1994}. However, we observe that the FZ and CVD samples show a very similar decay of the amplitude with temperature. The small differences may be sample-specific or originate from slight variations in the experimental parameters. 

We now turn to the temperature range from 180~K to 20~K. In this regime, the fluorescence is dominated by radiative decays. Still, one expects a temperature dependence of the fluorescence as a function of the excitation wavelength that follows the different population of the CF levels in the ground-state manifold. This can be seen in the PLE spectra in Figure~\ref{fig2}C. We used only the FZ sample in the following measurements because no sample dependence is expected in these measurements, as the CF levels are independent of the host material \cite{gritsch_narrow_2022}. Seven ground state transitions for each of the sites A and B are located between 1475~nm and 1538~nm. Because of the stronger overall signal, we insert the long-pass filter to eliminate the excitation laser and improve the signal-to-noise ratio. In addition, this leads to a stronger temperature dependence of the signal: With lower temperature, the population in the ground and excited manifolds is shifted towards the lowest level. This leads to a larger absorption at ground state transitions and a shift of the mean emission frequency to longer wavelengths, which increases the fluorescence signal transmitted by the long-pass filter. In addition, the narrowing of the transition peaks leads to an increase in their height.

The combination of these effects leads to a strong decay of the amplitude with temperature for a fixed wavelength. As an example, this is shown in Figure~\ref{fig2}D (red circles) for the $Z_1$-$Y_2$ transition of site A. To approximate the change of this observable with temperature, we fit the behavior phenomenologically with two exponential functions for the ranges of 18~K to 98~K and 103~K to 174~K (solid lines). While the measurement for a single excitation wavelength leads to a larger slope in temperature, averaging over several ground state transitions increases the size of our data set and thus the signal-to-noise ratio. Therefore, an average from 1505~nm to 1525~nm is plotted for comparison in Figure~\ref{fig2}D (red diamonds). Again, a phenomenological exponential fit estimates the rate of amplitude change with temperature.

\begin{figure}[ht]
\includegraphics[width=\columnwidth]{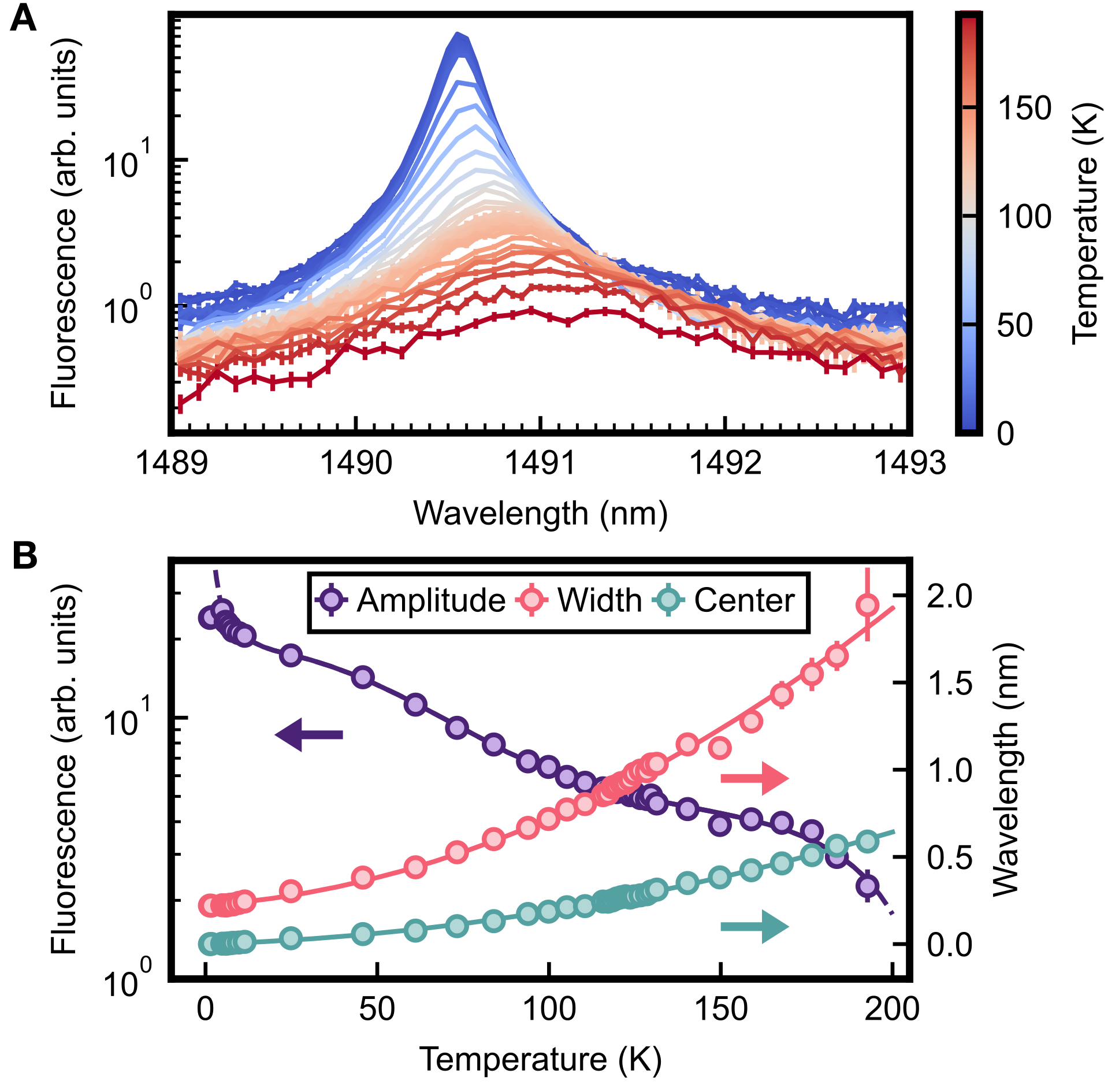}
\caption{Temperature-dependent study of the isolated $Z_1$-$Y_6$ transition of site A. (A) Photoluminescence excitation spectrum for various temperatures indicated by the color bar. (B) Lorentzian fits to the spectra are used to determine the amplitude (purple circles), full-width-half-maximum ("width", pink circles), and relative shift of the center wavelength ("center", teal circles). Arrows indicate the corresponding y-axis. The solid lines are phenomenological fits to extract an estimate of the slope. A spline fit is used for the amplitude, a polynomial of second order for width and center. Error bars in (A) correspond to the standard error of at least 400 averages and in (B) to estimated fit uncertainties based on the standard error of the data.
\label{fig3}}
\end{figure}

\subsection{Thermometry on an isolated transition\label{IsoPeak}}

As can be seen in the plot of the spectra in Figure~\ref{fig2}C, the signal obtained on many transitions overlaps with neighboring lines, which makes it difficult to study their isolated behavior. An exception is the well-isolated $Z_1 \rightarrow Y_6$ transition of site A around $\lambda^A_{Z_1Y_6} \simeq \SI{1490.6}{\nano\meter}$. In Figure~\ref{fig3}A, we plot the isolated fluorescence signal of this transition for temperatures in the range of 1.2~K to 209~K. 
By fitting a Lorentzian lineshape with a linear background, we can extract the amplitude, full-width-half-maximum (FWHM), and center of this peak. The results are shown in Figure~\ref{fig3}B. Several thermometric parameters can be extracted from this measurement: First, the amplitude (purple circles, left y-axis) that decreases with temperature because of the reduction of the population in $Z_1$; second, the increase of the FWHM (pink circles, right y-axis) caused by phononic relaxation; finally, also the center (teal circles, right y-axis) shifts to longer wavelengths. In the figure, this shift is referenced to the value at the coldest measured temperature, $\lambda^A_{Z_1Y_6}(T)-\lambda^A_{Z_1Y_6}(T_0)$. 

As can be seen, the amplitude of the peak changes by almost two orders of magnitude when going from cryogenic temperature to $\SI{200}{\kelvin}$, which allows for accurate temperature measurements in this wide range. However, in the regime of a few Kelvin, which is particularly interesting in quantum applications, the signal levels off, since the second level in the ground manifold $Z_2$ for site~A is gapped by an energy of  $k_B \times 126\ \mathrm{K}$, where $k_B$ is the Boltzmann constant. Thus, its population decreases below 10\% for a temperature under 30~K. Similarly, the phononic broadening of the absorption peak for ground state transitions starts to saturate. These observations reduce the sensitivity of the isolated-peak technique for temperatures below 10~K.

\subsection{Boltzmann thermometry using the spin levels\label{Boltz}}

The limitations imposed by the CF structure of the dopants can be overcome by applying a magnetic field, such that the Zeeman effect splits the two effective spin levels of each Kramers manifold in proportion to the field. Thus, the relative thermal population of the two spin states can be probed; this technique is called ratiometric or Boltzmann thermometry \cite{zeman_boltzman_2024}. At low temperature, only the two spin states $\left | \uparrow \right \rangle$ and $\left | \downarrow \right \rangle$ of the $Z_1$ level will be occupied. Under the assumption of equal transition strengths for the spin-preserving lines, the fluorescence $A_\uparrow$ and $A_\downarrow$ after selectively exciting these transitions allows for a calibration-free determination of the Boltzmann factor  \cite{zeman_boltzman_2024} using: 

\begin{equation}   \label{eq:T_boltzmann}
    T_r = -\frac{\Delta_g}{k_B \ln\left(A_\uparrow/A_\downarrow\right)},
\end{equation}

Here, $\Delta_g$ is the energy splitting of the spin levels in the chosen magnetic field configuration that can be measured independently \cite{holzapfel_characterization_2024}. It can be shown that for a given splitting $\Delta_g$, the ratiometric method has an optimum sensitivity range \cite{suta_theoretical_2020},
\begin{equation} \label{eq:optimal_T}
    \frac{\Delta_g}{k_B\left(2+\sqrt{2}\right)} < T < \frac{\Delta_g}{k_B 2}.
\end{equation}
Below and above this range, $T_r$ quickly saturates as either the population in the energetically higher spin state or the population change vanishes, such that the signal is governed by background and noise. 

For Er:Si, one finds effective g-factors of up to $g_\mathrm{eff}=\frac{h}{\mu_B}\times 230\ \mathrm{GHz/T}$, where $\mu_B$ is the Bohr magneton and $h$ is the Planck constant \cite{holzapfel_characterization_2024}. Given an inhomogeneous broadening of about 1 GHz \cite{gritsch_narrow_2022}, ratiometric thermometry will be most effective in the range below 10~K when applying magnetic field strengths exceeding approximately $\SI{1}{\tesla}$. 

However, in this regime, a potential limitation of the method is that the thermalization of the spin state can be impeded by a long spin-lattice relaxation time, which can exceed seconds in Er:Si \cite{gritsch_optical_2024}. Thus, care has to be taken that the population is distributed according to the equilibrium statistics. This is facilitated by operating at higher magnetic fields, as spin-lattice relaxation via the direct process increases proportional to the fifth power of the field amplitude \cite{shrivastava_theory_1983}. Similarly, the thermalization is improved at higher temperatures, where the Orbach process leads to an exponential reduction of the relaxation time \cite{shrivastava_theory_1983}.

Another difficulty can arise from the isotope $^{167}\text{Er}$ that is present in our samples (albeit it could be avoided by isotope-selective implantation). The nuclear spin of $^{167}\text{Er}$ can be very long-lived \cite{rancic_coherence_2018} at high magnetic fields. Thus, optical pumping of the nuclear spin state can lead to spectral hole burning and reduce the signal in our measurement. To avoid this, we modulate the laser frequency within a range of a few GHz such that it excites the whole inhomogeneously broadened ensemble independent of the nuclear spin state.  

\begin{figure}[ht]
\includegraphics[width=\columnwidth]{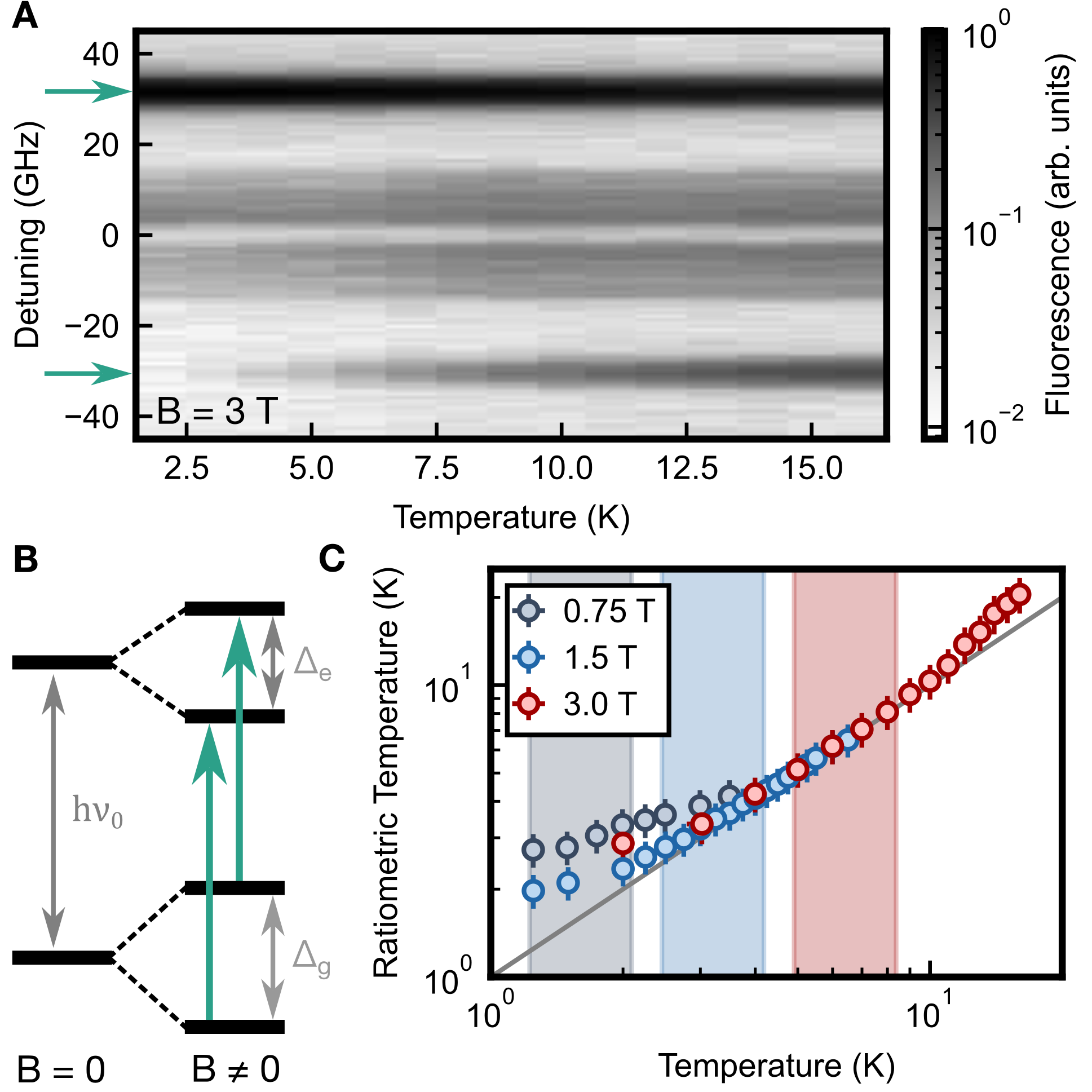}
\caption{Boltzmann thermometry. (A) The resonant photoluminescence spectra for site A show a clear temperature dependence at a magnetic field strength of 3~T. (B) Because of the difference of the Zeeman splittings of ground and excited state ($\Delta_g, \Delta_e$), the $Z_1 \rightarrow Y_1$ transition (at energy $h\nu_0$) splits symmetrically into several spin-preserving transitions. A pair of these peaks (teal arrows in A and B) is considered for thermometry. (C) The ratio of the peak amplitudes reflects the thermal population of the spin states according to the Boltzmann factor. The temperature extracted using this ratiometric technique deviates from the set value at low temperatures, where the increased spin lifetime impedes thermalization, and at high temperature, where higher-lying CF levels start to be populated. In between, the extracted temperature matches well with the expectation (grey line). Shaded areas depict the optimal sensitivity range according to the Zeeman splittings. Error bars in panel (C) are given by the statistical standard error of 750 averages and the uncertainty in the effective g-factors.
\label{fig4}}
\end{figure}

With this, we implement Boltzmann thermometry by applying a magnetic field strength of up to $B=3\ \mathrm{T}$ in the direction of the waveguide, along a [110] crystalline axis. The resulting temperature-dependent PLE spectra are shown in Figure~\ref{fig4}A. It can be seen that the spin-preserving $Z_1 \rightarrow Y_1$ transitions (schematically shown in Figure~\ref{fig4}B) can be resolved for several magnetic subclasses of site A. For the given magnetic field orientation, we can use the previously determined g-tensor for this site to explain all appearing peaks and their splitting \cite{holzapfel_characterization_2024}. The peaks originating from one of these classes (marked by two arrows) are clearly separate from the others and have a higher brightness due to a larger degeneracy, and will thus be used in the following.  

The Zeeman splitting $\Delta_g$ of this class gives an effective g-factor of $g_\mathrm{eff}=\frac{h}{\mu_B}\times 116(13)\ \mathrm{GHz/T}$. The peak amplitudes $A_\downarrow$ and $A_\uparrow$ are extracted using Gaussian fits. This eliminates the effect of a small offset of the PL signal caused by background and detector dark counts. With this, under the assumption of equal transitions strengths for the spin-preserving lines \cite{zeman_boltzman_2024}, Equation \eqref{eq:T_boltzmann} can be used for a calibration-free temperature measurement. The extracted values are compared to the temperature of the external sensor in Figure~\ref{fig4}C for three magnetic field strengths, together with the optimum sensitivity region according to Equation \eqref{eq:optimal_T} (shaded colors). One observes excellent agreement with the expected temperature (grey diagonal line) in the optimum range at temperatures between 3~K and 10~K for the measurements at 1.5~T and 3~T. 

However, if the magnetic field is decreased to resolve lower temperatures, the increased spin lifetime \cite{shrivastava_theory_1983} prevents that thermalization is faster than the spin pumping that results from the optical probing. This leads to a systematic shift of the $B=0.75~\mathrm{T}$ points (grey circles) that may be avoided by slower pulse repetition rates at the price of an increased measurement duration. Furthermore, a clear deviation also arises well above the optimum sensitivity window for $B=3~\mathrm{T}$ data (red circles) at about 10~K. Here, the timescale of phonon-induced transitions between the spin states (via an Orbach process mediated by the $Z_2$ level) will be faster than the optical lifetime \cite{gritsch_narrow_2022} and the excitation pulse duration. Thus, the fluorescence signal will no longer be proportional to the initial population of the probed spin state, leading to deviations in the inferred temperature. Potentially, this could be resolved using shorter excitation pulses. In conclusion, the thermometric method needs no further calibration only in a certain range; however, a larger range can be used with a calibration of the systematic shifts.

\subsection{Comparison of the different thermometry methods}
In this section, we assess and compare the performance of the presented thermometric probes $y(T)$ using the relative thermal sensitivity $S_r$ and temperature precision $\delta T$ \cite{brites_spotlight_2023},
\begin{equation}
    S_r = \frac{1}{y(T)}\left|\frac{\partial y}{\partial T}\right|,\ \delta T = \pm \frac{1}{S_r}\left| \frac{\delta y}{y} \right|,
\label{eq:thermal}
\end{equation}
where $\delta y$ is the statistical uncertainty of the thermometric probe. 
For a simplified estimation of the slope of the thermometric probes, we use the phenomenological fits that provide an interpolation of the slope in $y$. 

\begin{figure}[ht!]
\includegraphics[width=\columnwidth]{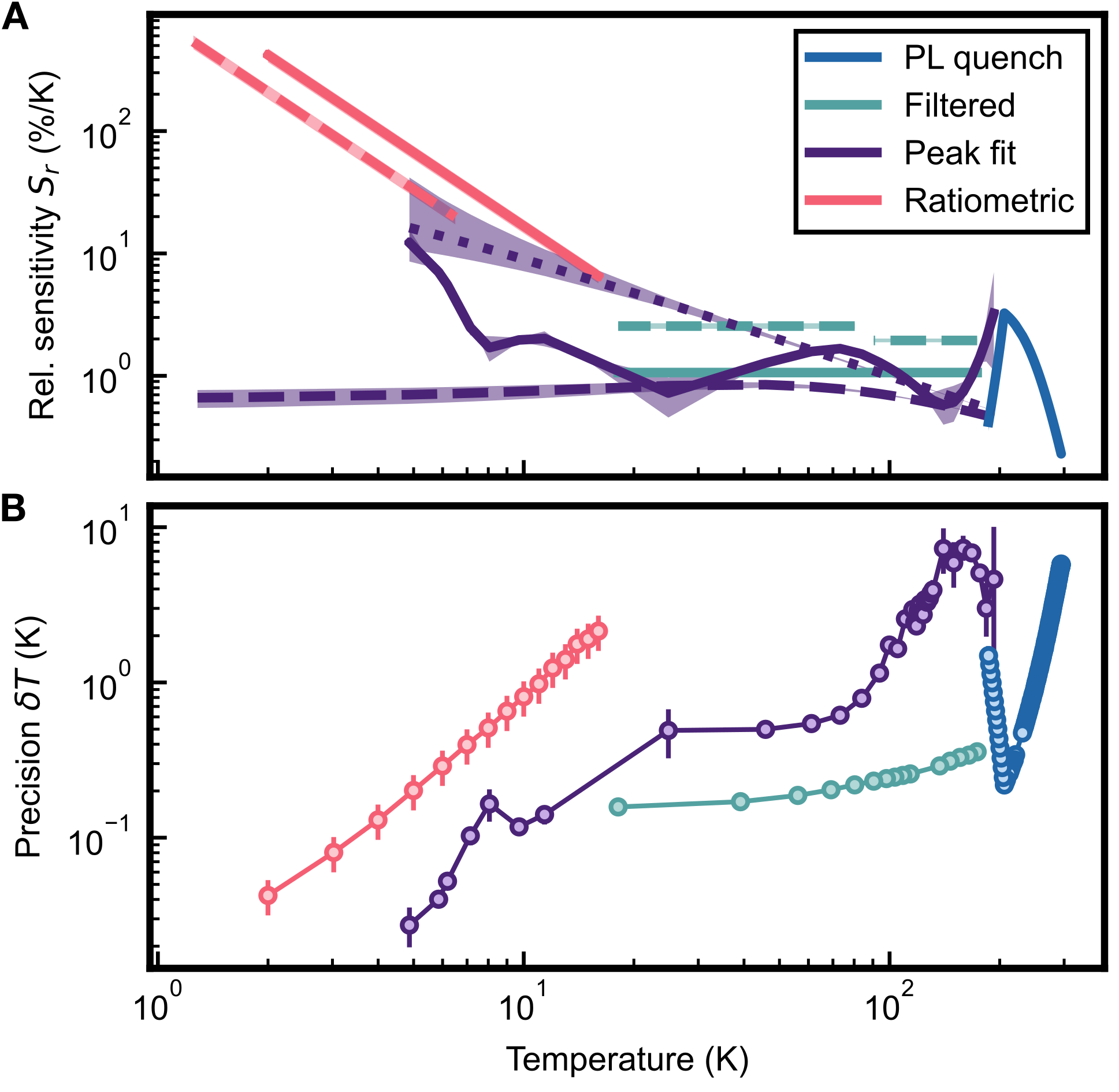}
\caption{Assessment and comparison of the thermometric parameters. The relative thermal sensitivity $S_r$ (A) and the temperature precision $\delta T$ (B) defined by Equation~\eqref{eq:thermal} are shown for the thermal probes introduced in subsections~\ref{PLquench}-\ref{Boltz}. $S_r$ is obtained using the phenomenological fits for their valid temperature regime. The blue data ("PL quench") shows wavelength-averaged amplitude response for the CVD (solid, circles) samples. The teal data ("Filtered") compares wavelength-averaged (solid, circles) and single-wavelength (dashed) measurements on the FZ sample with long-pass filter. The purple data ("Peak fit") depicts the amplitude (solid, circles), wavelength shift of the center (dotted) and full-width-half-maximum (dashed) obtained from a Lorentzian fit to the isolated $Z_1$-$Y_6$ transition response of site A. The pink data ("Ratiometric") shows the results of Boltzmann thermometry on split Kramers doublets for magnetic field strengths of 3~T (solid, circles) and 1.5~T (dashed). In panel B, we present the precisions $\delta T$ of a subset of the methods presented in A. Teal data shows wavelength-averaged, purple the fitted amplitude, and pink data taken at \SI{3}{\tesla}. The shaded area in (A) and error bars in (B) represent the uncertainty of the fits.
\label{fig5}}
\end{figure}

In Figure~\ref{fig5}A, we plot $S_r$ for the thermometric probes introduced in subsections~\ref{PLquench}-\ref{Boltz}. This parameter is then used to weigh the relative uncertainties of the probe data, which gives the temperature precision $\delta T$ in Figure~\ref{fig5}B. We start by analyzing the quench regime. The blue data ("PL quench") denotes an average of the PLE spectrum (see Figure~\ref{fig2}A and B) of the CVD sample between 1520~nm and 1550~nm using an exponential with offset to fit the slope. As mentioned, the FZ sample gives comparable results (not shown). The time resolution of this probe with a total of $7.5\cdot 10^4$ data points per individual probe point is 5~minutes. The rapid suppression of the signal deteriorates the sensitivity from a few percent per Kelvin to a few permille per Kelvin when approaching room temperature.

We now turn to the dataset of Figure~\ref{fig2}C and D that covers intermediate temperatures because the long-pass filter is inserted to increase the thermal sensitivity. The thermometry performance is shown in teal color ("Filtered") in Figure~\ref{fig5} that includes a comparison between averaging the spectrum over a broader range of wavelengths (1505~nm-1525~nm, solid line) and data taken at a single wavelength (1519~nm, dashed line). The sensitivity is significantly better for the single-wavelength measurement. Derived from the same raw data, the single-wavelength data has an integration time of \SI{0.5}{\second} and the averaged data uses $10^5$ points, leading to an integration time of 7~minutes. We only plot the latter in Figure~\ref{fig5}B to provide a fair comparison to the other methods shown that are taken with similar integration times. 

Next, we discuss the data extracted from the isolated $Z_1 \rightarrow Y_2$ transition of site A (see Figure~\ref{fig3}). It is shown in purple color and includes the probes based on the peak center shift (dotted line), amplitude (solid line), and FWHM (dashed line). The fit is done on overall $4\cdot 10^4$ data points, yielding an integration time of 3~minutes. The FWHM and shift data are not directly dependent on the excitation parameters and, thus, are less prone to power fluctuations. Still, the amplitude data shows the best precision over the full temperature range and is thus depicted in Figure~\ref{fig5}B. 

Finally, the method of measuring the population imbalance of the Kramers doublet ("Ratiometric") is shown in pink for the two magnetic field strengths $B=3\ \mathrm{T}$ (solid lines) and $B=1.5\ \mathrm{T}$ (dashed line). As mentioned, this method exhibits a calibration-free temperature measurement, but is limited to low temperatures. Still, it provides the highest sensitivity range. Its precision is similar for both magnetic field strengths and is represented in Figure~\ref{fig5}B by the $B=3\ \mathrm{T}$ data achieving values below 100~mK. For the extraction of the population ratio, we use $1.6 \times 10^5$ data points per measured temperature, resulting in an integration time of about 11~minutes. 

Across the methods presented in Fig.~\ref{fig5}A, we extract a maximum of the relative sensitivity of \SI{420(50)}{\percent\per\kelvin} for the ratiometric method at the coldest temperature. The sensitivity then decreases towards room temperature, where we extract a value of \SI{0.22(4)}{\percent\per\kelvin}. The temperature precisions plotted in Fig.~\ref{fig5}B are extracted with integration times between three and eleven minutes, with a minimum value of \SI{0.04(1)}{\kelvin} for the fitted amplitude method at low temperatures and a maximum value of \SI{6(1)}{\kelvin} at room temperature. 

\section{Conclusion}
We have measured temperature-dependent resonant photoluminescence spectra for an erbium ensemble integrated into a nanophotonic silicon waveguide in a wide range of temperatures. We find that different thermometric probes give the best precision depending on the temperature regime: From 300~K to 200~K, the observed quench of the fluorescence renders amplitude-based probes as the most sensitive. However, the small signal strength requires longer integration times to achieve high precision close to room temperature, and heating caused by the large non-radiative decay might distort the measurement in thermally isolated samples.

At lower temperature, the crystal field structure of erbium, with a total of 56 transitions spread over more than 100~nm, leads to a strongly temperature-dependent mean absorption and fluorescence wavelength. This is exploited using a spectrally selective filter in the detection of the amplitude-based probes on ground-state transitions in the temperature regime 200~K to 20~K. Optimization of the spectral shape of the excitation source and the detection filter promises further increased precision with better time resolution. Since non-radiative decays are suppressed in this temperature regime, probe-induced heating can occur via Stokes processes and background absorption. While increased losses for erbium-implanted waveguides \cite{rinner_erbium_2023} suggest higher background absorption, a quantitative study of the induced heating is still missing. 

Finally, in the regime between 10~K and 2~K, the best results are achieved with a ratiometric temperature probe that measures the population difference in the Kramers doublet when split by a magnetic field. This provides a calibration-free method to extract the temperature. The extension to colder temperatures is currently limited by the increasing spin lifetimes at low magnetic field strengths and temperatures leading to optical spin pumping when probing the population. 

The precision achieved in our first demonstration of temperature measurements based on emitters in silicon nanophotonic waveguides is already sufficient for many applications in classical and quantum technology. Practical devices may use the studied waveguide geometry or different types of nanophotonic structures. The technique is fully compatible with foundry-based silicon nanofabrication \cite{rinner_erbium_2023}, and we expect that the technique can also be directly transferred to other crystalline nanophotonic platforms. The performance can be further improved in the future by better off-chip coupling, optimized sensing sequences, and a higher integration yield into the desired sites A and B \cite{gritsch_narrow_2022}. Excitation and collection efficiencies can be further boosted by embedding the emitters into photonic crystal waveguides \cite{arcari_near-unity_2014} or cavities \cite{gritsch_purcell_2023}. Combining this with spatially selective implantation promises precise thermometry from ambient to cryogenic temperatures with a spatial resolution down to a few nanometers. This opens the door for a large number of experiments in which thermalization within nanophotonic components can be studied with a previously inaccessible spatial resolution.

\begin{acknowledgments}
This project received funding from the Deutsche Forschungsgemeinschaft (DFG, German Research Foundation) under the German Universities Excellence Initiative - EXC-2111 - 390814868, from the German Federal Ministry of Education and Research (BMBF) via the grant agreement No 16KISQ046, and from the Munich Quantum Valley, which is supported by the Bavarian state government with funds from the Hightech Agenda Bayern Plus.
The datasets generated during and/or analyzed during the current study are available in the mediaTUM repository, \href{https://doi.org/10.14459/2025mp1768103}{https://doi.org/10.14459/2025mp1768103}.
\end{acknowledgments}


\end{document}